\documentclass[%
floatfix,
 reprint,
 amsmath,amssymb,
 aps,
]{revtex4-2}
\usepackage[top=2cm,bottom=2cm,left=1.5cm,right=1.5cm,columnsep=0.7cm]{geometry}

\usepackage[T1]{fontenc}
\usepackage[utf8]{inputenc}
\usepackage[english]{babel}

\usepackage{amsfonts,amssymb,amsmath,array}
\usepackage{bm}
\usepackage{dcolumn}
\usepackage[output-decimal-marker={.}]{siunitx} 

\usepackage[caption=false]{subfig}
\usepackage{comment}
\usepackage{graphicx} 
\usepackage{listings} 
\usepackage[usenames]{color} 
\usepackage{subfig}
\usepackage{ulem}
\usepackage{microtype}
\usepackage{color}
\usepackage{float}

\usepackage{hyperref}
\hypersetup{
    colorlinks=true,
    linkcolor={blue}, 
    filecolor=magenta,   
    citecolor={blue}, 
    urlcolor=blue, 
}

\newcommand{\br}{{\bf r}}

\newcommand{\bF}{{\bf F}}
\newcommand{\bn}{{\bf n}}

\begin{document}

\preprint{APS/123-QED}

\title{Liquid-Hexatic-Solid phases in active and 
passive Brownian particles determined by
stochastic birth and death events}

\author{Alejandro Almod\'ovar}
\email{almodovar@ifisc.uib-csic.es}
\author{Tobias Galla}
\author{Crist\'obal L\'opez}
\affiliation{IFISC, Instituto de F\'isica Interdisciplinar
y Sistemas Complejos (CSIC-UIB), Campus Universitat de les
Illes Balears, E-07122 Palma de Mallorca, Spain}

\begin{abstract}
We study the effects of stochastic birth and death processes on the structural phases of systems of active and passive Brownian particles subject to volume exclusion. 
The total number of particles in the system is a fluctuating quantity, determined by the birth and death parameters, and on the activity of the particles. As the birth and death parameters are varied we find liquid, hexatic and solid phases. For passive particles these phases are found to be spatially homogeneous. 
For active particles motility-induced phase separation (co-existing hexatic and liquid phases) occurs for large activity and sufficiently small birth 
rates. We also observe a re-entrant transition to the hexatic phase when the birth rate is increased. This results from a balance of an increasing number of particles filling the system,  and a larger number of defects resulting from the birth and death dynamics. 
\end{abstract}

\maketitle

\section{Introduction}

Active matter \cite{Marchetti2013,Bechinger2016} has been intensively investigated in the last decades partly due to its many applications to living systems \cite{Hakim2017}.  Examples include flocks of birds \cite{Cavagna2014}, schools of fish \cite{Larsson2012,Shaw1978}, the swarming of bacteria \cite{Dombrowski2004}, and the migration of cancer cells \cite{Chepizhko2018}. Dense particle configurations are particularly relevant for biological processes such as wound healing \cite{XTrepat2020,XTrepat2021,Angelini2011} and tissue formation \cite{Drongelen2018}. The interplay between motility, interactions, biochemical dynamics and other effects induce interesting collective phenomena in these systems \cite{Dombrowski2004,Szabo2006,Angelini2010,parrish1997}. 

A key model in this area is that of active Brownian particles (ABP). This model describes a system of self-propelled particles which interact only through mutual volume exclusion \cite{Romanczuk2012,tenHagen2011,Fily2012}. In two dimensions, the nonequilibrium phase diagram at  high densities was established in \cite{digregorio2018}. Broadly, it shows the same types of ordering as also seen in equilibrium two-dimensional melting of hard disks \cite{Bernard2011,Kapfer}, namely liquid,  hexatic and solid phases. Activity in the system of Brownian particles shifts transition lines towards higher densities and motility-induced phase separation (MIPS) becomes possible \cite{Cates2015,Caprini2020_2,Buttinoni2013}. The basic structural phase 
diagram becomes more complex if one considers velocity correlations in space \cite{caprini2020}.

In this work we add a simple but general biological ingredient. We allow the stochastic birth and death of particles, and focus on the regime in which the time scales of the birth and death processes are comparable to those of the movement of particles. Thus we aim to introduce a minimal model to describe the spatial ordering for example of living microorganisms, capturing some of the effects of their size, movement and stochastic population dynamics. Our analysis encompasses systems with active motion as well as  so-called passive Brownian particles (PBP). In contrast with conventional models of ABP or PBP,  the packing fraction in our setup
is a time-varying quantity (due to the birth and death dynamics), and fluctuates  around a steady-state value at long times.

Our focus is on the structural phases of this system. We restrict the analysis to the case in which the birth rate is larger than the death rate,
so that the system reaches a state of high density eventually
 (if the death rate is higher than the birth rate the system typically ends up being empty). 
Some of the questions we seek to address are the following:
 Does the demographic dynamics determine the spatial distribution of the particles of
both systems, and if so, how? What is the interplay between  self-propulsion and population dynamics, in particular, concerning the coexistence
of different phases (e.g. MIPS)? What is the role of
 defects in this context? 
Addressing these questions may provide 
further insight, at a qualitative level,
into biological phenomena such as 
wound healing and tissue formation, which have been 
studied before from the 
perspective of  active matter \cite{Zapperi,Sepulveda}.

We note that even the case of passive particles constitutes a nonequilibrium system in our model. This is due to the demographic dynamics. In the passive case 
 we observe the same uniform phases  as in equilibrium two-dimensional melting (liquid, hexatic and solid) \cite{Farid1981,Kosterlitz1973,Nelson1979,Halperin1978,Young1979}. Importantly, which one of these phases is realised at fixed temperature (or diffusivity) is determined by the birth and death rates. As a central part of our work we provide a full phase diagram.

When the particles are active, we observe a MIPS-like phase in the system with birth and death dynamics, similar to what is observed in a conventional ABP system without birth and death. We provide a phase diagram in the space defined by the activity parameter and the birth rate for a 
fixed value of the death rate. We also discuss differences compared to a conventional ABP system.

The remainder of this paper is organised as follows. In Sec. \ref{Sec:model} we define the model of passive or active particles, undergoing motion and birth-death dynamics. We also describe the numerical method we use to simulate the dynamics. Sec. \ref{Sec:results} contains our main results. In Sec. \ref{Sec:summary} we 
present a summary and discussion.

\section{Model and numerical algorithm}
\label{Sec:model}

The model consists of two components. One is the interaction and movement of particles. This is as in the conventional setup of \cite{digregorio2018}, where particles interact via volume exclusion and follow a standard Langevin dynamics. The second component describes the birth and death of particles. We here use exponential 
processes to trigger deaths and potential births. This is simulated most conveniently using the celebrated Gillespie algorithm \cite{Gillespie2007}. We now present details of the setup and of the simulation method.

\subsection{Passive and active Brownian particles}

We consider a two-dimensional system of $N$ interacting disks in the overdamped limit (i.e., we neglect inertia). The particles move in a square domain of size $L\times L$, with periodic boundary conditions. We also discard any hydrodynamical interaction. The central position of every disk evolves in time according to the Langevin equation,
\begin{equation}
  \dot{\br_i}= \frac{1}{\gamma} {\bF}_i + \frac{1}{\gamma} {\bF}^{act}_i + \sqrt{2D} \boldsymbol{\zeta}_i (t),
\label{eq:langevin}
\end{equation}
where $\gamma$ is the friction coefficient. The $\{\boldsymbol{\zeta}_i\}$ are independent Gaussian noise vectors satisfying
$\langle \boldsymbol{\zeta}_i \rangle= 0$, $\langle {\zeta}_{i,a} (t) { \zeta}_{j,b} (t') \rangle = \delta_{i j} \delta_{a b}\delta (t-t')$ (here, $a$ and $b$ describe the entries 
of the two-component vectors $\boldsymbol{\zeta}_i$ and $\boldsymbol{\zeta}_j$). The parameter $D>0$ is related to temperature via Einstein's relation $D=k_B T/ \gamma$. The interaction  force is assumed to derive from a pairwise interaction potential $U$, i.e., $\bF_i= -\nabla_i \sum_{j\neq i} U(|\br_i -\br_j|)$. 
We follow \cite{digregorio2018,caprini2020} and choose a truncated Lennard-Jones potential to describe finite-size particles. 
 Specifically, we set $U(r) = 4 \varepsilon [(\frac{\sigma}{r})^{12} - (\frac{\sigma}{r})^6] + \varepsilon$, 
 if $r<\sigma_d \equiv 2^{1/6} \sigma$, and $U(r)=0$ if $r > \sigma_d$. 
The cut-off at $r=2^{1/6}\sigma$ is to remove the attractive part of the potential 
to mimic hard sphere repulsion.
The quantity $\sigma>0$ represents the particle diameter,
and  $\varepsilon$ is an energy scale. The additive 
constant $\varepsilon$ for $r<\sigma_d$ is irrelevant for the dynamics, its purpose is to make the potential continuous. The resulting potential is known as the Weeks-Chandler-Andersen (WCA) potential \cite{WCA}. 

Our analysis focuses on two distinct scenarios. In the first one, the particles are passive, i.e., $\bF^{act}_i = 0$. 
We refer to this as a system of passive Brownian particles. In the second scenario the particles are active (i.e., self-propelled). For such active Brownian particles, we use self-propulsion forces $\bF^{act}_i = v_0 \bn[\theta_i (t)]$, of constant modulus $v_0$ and with direction (in the plane) given by the unit vector 
$\bn (\theta_i) = ( \cos \theta_i, \sin \theta_i )$. The angle $\theta_i$ for particle $i$ performs diffusive motion, $\dot{\theta}_i (t) = \sqrt{2 D_r} \eta_i (t)$, where $D_r$ is the so-called `rotational diffusion coefficient' \cite{Saragosti2012,Jain2017}. We take these to be Gaussian with zero mean, and $\langle \eta_i (t) \eta_j (t')\rangle = \delta_{ij} \delta (t - t')$. 
We treat $D$ and $D_r$ as independent parameters.

\subsection{Birth and death dynamics}

Particles may randomly reproduce (self-replicate) or die, so that the total number of particles, $N(t)$, changes with time. Death events occur through a Poisson process with constant per capita removal rate $\delta$. The corresponding particle is then simply removed from the system. 

Birth events are more elaborate. In previous studies particles had no spatial extension, and offspring were located at the same place as 
the parent \cite{EHGCLopez2004,FRamos2008,Heinsalu2010,Young2001}. In \cite{Khalil2017} the particles had a finite radius, but the model operated at coarse-grained level of the density of particles (per area). In the present work particles have a fixed diameter $\sigma$. This combined with volume exclusion complicates the implementation of the birth process. 

Potential birth events in our model are triggered with per capita rate $\beta$. That is to say an existing particle is selected for potential reproduction. However, the birth event can only go ahead if there is room for the offspring. More precisely, we assume that new particles are placed at a distance  $\sigma$ from the parent. In two dimensions this defines a continuous ring of possible positions for the centre of the offspring. For practical implementation, we restrict this to $4n_\varphi$ discrete positions, placed at equal angular separation on the ring (we choose $n_\varphi=500$). Thus, the points at which the offspring can potentially be located are at distance $\sigma$ from the centre of the parent particle, and at angular positions $0,\frac{\pi}{2n_\varphi},\frac{\pi}{n_\varphi},...,\frac{\pi(4n_\varphi-1)}{2n_\varphi} $. In the first instance one of these positions is selected at random. If placing the offspring at that position does not lead to an overlap with any existing particle, then the birth event completes. If there is overlap, then a new potential position is chosen at random from the remaining $4n_\varphi-1$ possible options. This procedure continues until either the offspring is successfully placed, or until all $4n_\varphi$ positions have been exhausted. In this latter case the reproduction event does not occur, and the simulation continues. 

While $\delta$ is the fixed per capita death rate, we note that $\beta$ is not the actual birth rate. Instead, it is the maximum possible per capita birth rate which would be realised if all proposed birth events complete. In practice, the typical time between two birth events in the population will be higher than $1/(N\beta)$, because some reproduction events that are triggered may not complete. This effective birth rate is therefore a stochastic observable, dependent on the positions of the particles. The larger the density of particles is around a focal particle, the lower the true birth rate becomes for this individual. The number of particles in the system increases when the average birth rate is higher than the death rate. Conversely, if the density is too high, death events dominate, and the number of particles decreases. At long times, the particle number reaches a stationary distribution (see Fig.~\ref{Fig:Nvst}). In this stationary state the mean effective birth rate is equal to the death rate.

\subsection{Algorithm}

To simulate the system we use a combination of the well-known Gillespie algorithm for the birth and death dynamics \cite{Gillespie2007}, and an Euler-Maruyama integration of the Langevin equations for the motion of the particles. The latter requires the choice of a discretisation time step $\Delta t$, and careful interfacing with the continuous-time Gillespie algorithm is required. We now describe our simulation method in detail. Assuming time $t$ has been reached and that the system contains $N(t)$ particles at that time, the algorithm proceeds as follows:

\begin{enumerate}

\item Generate an exponentially distributed random number $\tau$ with parameter $N(t)\times (\delta+\beta)$. 

\item Update the positions of all particles by integrating the Langevin equations forward to time $t+\tau$. To do this, proceed as follows:

\begin{enumerate}
\item[(a)] If $\tau>dt$ carry out $k+1$ Euler-Maruyama steps to integrate Eqs.~(\ref{eq:langevin}), where $k$ is the largest integer smaller than $\tau/dt$ (i.e., $k$ is the number of steps of length $dt$ that fit into an interval of length $\tau$). The first $k$ if these Euler-Maruyama steps use time step $dt$. The remaining step is of length $\tau-k dt$. 
 \item[(b)] Or, if $\tau<dt$ carry out a single Euler-Maruyama iteration with time step $\tau$.
\end{enumerate}

\item Once the time $t+\tau$ has been reached, a possible birth or death event is carried out. More precisely, a death event occurs with probability $\delta/(\delta+\beta)$, or a birth event is proposed with complementary probability $\beta/(\delta+\beta)$. 

\begin{enumerate}
\item[3.1] In the case of a death event choose one of the $N(t)$ particles at random (with equal probabilities) and remove the particle from the system. Go to step 4.

\item[3.2] If a birth event is proposed, choose one particle at random (with equal probabilities) for possible reproduction. They follow the procedure described above in the text to decide if the birth event can go ahead. If it can, place the offspring at the designated position. If the birth event is not possible due to particle exclusion then no new particle is produced. Either way, go to step 4.
\end{enumerate}

\item The algorithm has now reached time $t+\tau$. The number of particles at that time, and their positions are known. Go to step 1.
\end{enumerate}

We fix the following parameters in the numerical simulations: $\gamma=1$, $k_B T = 0.05$,  $\epsilon=1$ and $\sigma=1$,  for both active and passive particles. In addition, for the active case we use $D_r = 1.0$, and  $v_0$ will be varied. The particles are in a two-dimensional box of length $L=70$ with periodic boundary conditions. The number of possible positions for a newborn disk is $4n_\varphi=2000$.

\section{Results}
\label{Sec:results}

\subsection{Passive particles}
\label{subsec:passive}
\subsubsection{Stationary packing fraction}
\begin{figure}[hbt!]
\begin{center}
\includegraphics[scale=0.37]{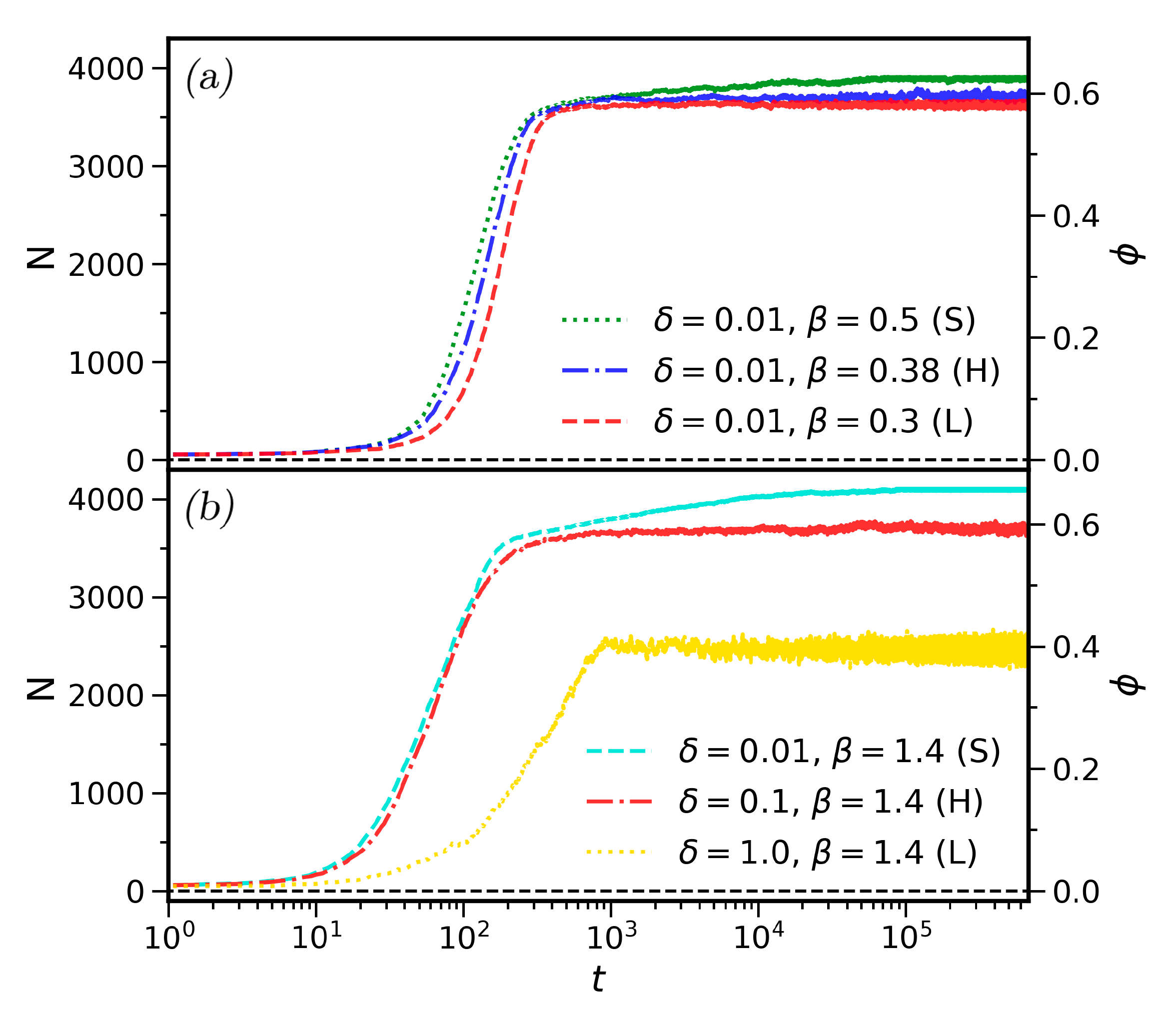}
    \caption{Number of particles as a function time for different values of the birth and death rates. (The corresponding packing fraction $\phi$ is indicated on 
the vertical axis on the right of each panel). 
(a) Fixed value of the death-rate $\delta$, and varying 
values of the birth parameter $\beta$.
(b) Fixed value of $\beta$, varying death rate $\delta$. 
The letters L, H or S indicate the phase the dynamics 
results in for the different parameter sets (liquid, hexatic or solid).}
    \label{Fig:Nvst}
\end{center}
\end{figure}

We first discuss the case of passive particles. We focus on the regime $\beta>\delta$ (birth more frequent than death), as the system otherwise evolves to an empty absorbing state. In Fig.~\ref{Fig:Nvst} we show how the number of particles in the system changes in time for different choices of the birth and death parameters $\beta$ and $\delta$. The data confirms that the number of particles at long times, or equivalently the packing fraction $\phi (t)=N(t)\pi (\sigma/2)^2/L^2$, strongly depends on the relative strength of the birth and death processes. For fixed death rate $\delta$ (upper plot in Fig.~\ref{Fig:Nvst}) the long-time particle number increases with the birth rate. 
 Also, if we fix $\beta$ (bottom plot), as the value of the death rate increases
 the number of particles diminishes, but fluctuations in the particle
 number increase.

One might expect a complete filling of the system of 
particles independently of the birth and death rates,
 but results show this is not so. Instead there 
 is a more complicated interplay between 
 motion and population dynamics. 
 For large reproduction rates, as soon as there is room, a new
particle is born.
For smaller birth parameter, this is not necessarily the case, mainly due to the motion of particles.

\begin{figure*}[t!]
\begin{center}
\includegraphics[scale=0.625]{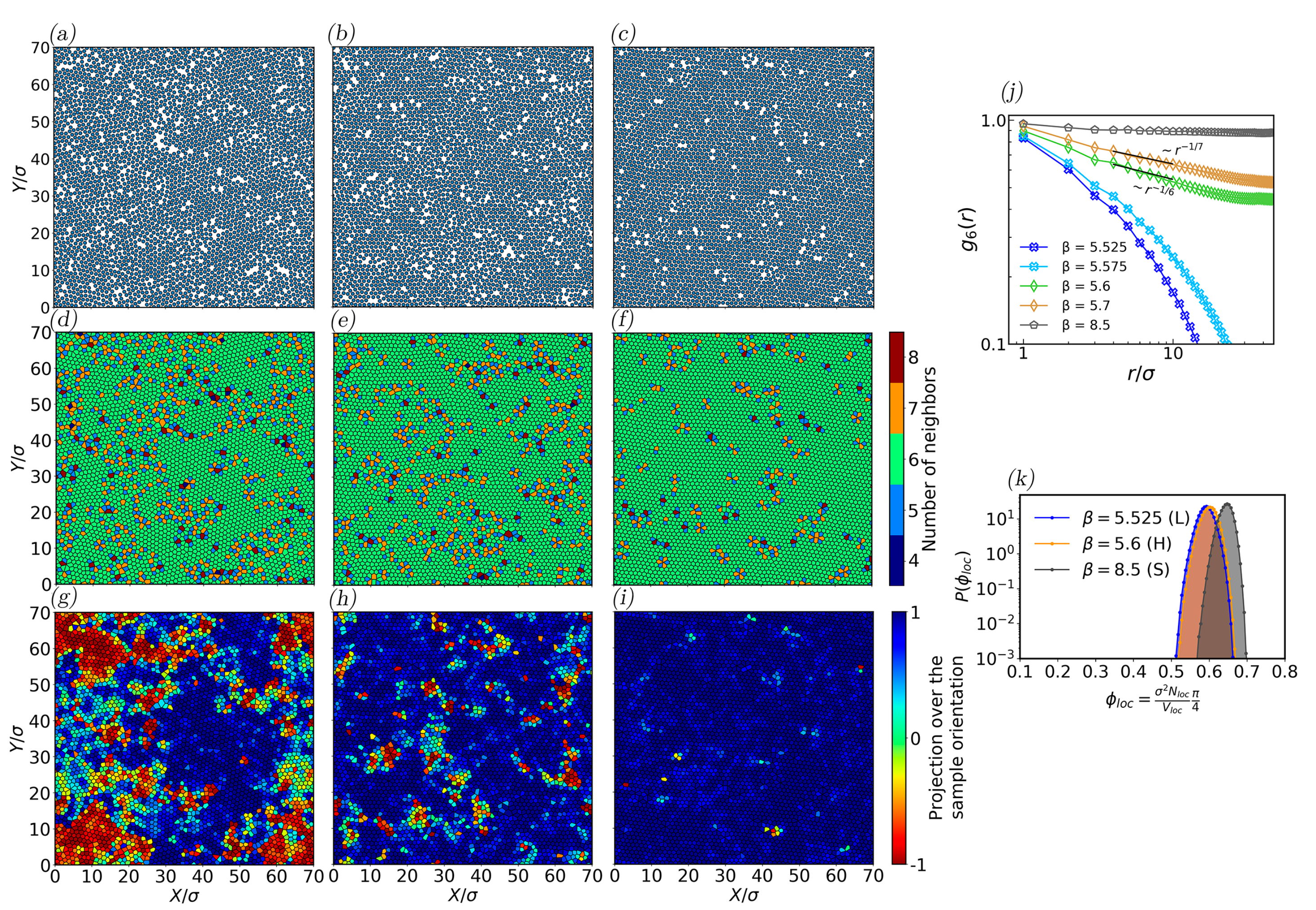}
    \caption{Structural properties of the system in the different phases. Panels in the first column [(a), (d), (g)] are in the liquid phase, those in the the second column [(b), (e), (h)] are in the the hexatic phase, and the panels  (c), (f), and (i) are in the solid phase. We have used $\delta=1.0$ in all plots, and
    three different values of $\beta$: $\beta = 5.525, 5.6, 8.5$ from left to right.
    Panels (a), (b), and (c) show snapshots of the system in real space at long times. Panels
    (d), (e), and (f) are the corresponding Voronoi tessellations, where the color indicated the number of neighbors for the different particles.
    Panels (g), (h), (i)  show again the Voronoi tessellation but the colors now represent the projection of local orientation on the direction of the sample orientation (see text).
    Panel (j) shows the correlation function $g_6(r)$ (see text) for different values of the birth parameter $\beta$. The lower two lines (crosses) are in the liquid phase, the upper line (pentagons) in the solid phase, and the remaining lines (diamonds) in the hexatic phase. Panel (k) shows the distribution, $P(\phi_{loc})$, of the local packing fraction for  $\beta = 5.525, 5.6, 8.5$.}
    \label{Fig:Plot}
\end{center}
\end{figure*}

\subsubsection{Characterisation of the different  phases}
In two-dimensional equilibrium melting the packing fraction (at a fixed temperature) determines whether the system is in a liquid, hexatic or solid state \cite{Bernard2011}. Thus, we expect that changing the parameters of the birth and death dynamics can induce transitions between similar phases. To confirm this, we show different spatial  configurations of the system in the steady state in Fig.~\ref{Fig:Plot}. These are for different choices of the birth and  death parameters, panels (a)-(c) show the arrangement of particles in space. Panel (a) is in the liquid phase, (b) in the hexatic phase, and (c) in the solid phase.
 
To confirm the structural nature of the different phases we use the orientational order parameter 
$\psi_6 (\br_i) = \sum_{j\in i} e^{6 i \alpha_{ij}} /N_i $ \cite{Gasser2009}, where $\br_i$ describes the position of the $i$-th particle. 
The sum is over the $N_i$  neighbors of particle $i$ (we write $j\in i$ if $j$ is a neighbor of $i$). In order to define neighborhood relations we first determine the Voronoi tessellation of the box. In other words, there is one Voronoi cell for each particle. The cell for particle $i$ consists of the points in space which are closer to particle $i$ than to any other particle. This is illustrated in Fig.~\ref{Fig:Plot} (d)--(f). Two particles are considered neighbors if their Voronoi cells are in contact \cite{Rycroft}. The quantity $\alpha_{ij}$ in the definition of $\psi_6(\br_i)$ is the angle with respect to the $x$-axis of the line joining the positions of particles $i$ and $j$. 

 The parameter $\psi_6(\br_i)$ quantifies how ordered the positions of the neighbors of particle $i$ are. For a perfect triangular lattice the quantities $6\alpha_{ij}$ are the same for all $j\in i$, and hence $|\psi_6(\br_i)|=1$. The projection $\langle\psi_6(\br_k) \boldsymbol{\hat{e}}_\Psi \rangle$ is shown as a color map in Fig.~\ref{Fig:Plot} (g)--(i), where $\boldsymbol{\hat{e}}_\Psi$ is the unit vector pointing in the direction of the sample orientation $\Psi=(1/N) \sum_k \psi_6(\br_k)$. Zones of the same color in those panels of the figure show domains with the same orientation of the spatial ordering, indicating long-range orientational correlations.

The defects emerging in the system are the standard ones \cite{Digregorio2022}, disclinations, dislocations and vacancies. The defects are identified in the Voronoi representations [Fig.~\ref{Fig:Plot} (d)--(f)] as particles with a number of neighbors different from six. In the liquid phase, Fig.~\ref{Fig:Plot}(d), mainly there are disclinations (one-particle defects) and dislocations (pairs of neighboring particles, with 5 and 7 neighbors respectively). No disclinations are found in the hexatic phase [see Fig.~\ref{Fig:Plot}(e)]. In the solid phase [Fig. \ref{Fig:Plot}(f)] there are only few defects, and nearly all of them are of the vacancy type \cite{Digregorio2022} (point defects resulting from the absence of a particle in an otherwise hexagonal packing). 
These holes are due to the population dynamics (in particular death), and they do not break the hexagonal symmetry as shown in Fig.~\ref{Fig:Plot}(i). Most of these holes get filled quite quickly. This type of defect is also present in the other phases although it is not easy to visualize because there are in fact many defects.

From the orientational order parameter, we define the correlation $g_6 (r)=\langle\psi_6 (\br_i) \psi_6^*(\br_j)\rangle /\langle|\psi_6(\br_j)|^2\rangle$ (where $|\br_i-\br_j|=r$). We average $g_6(r)$ over  time in the stationary state, results are shown in Fig.~\ref{Fig:Plot}(j) for different choices of the model parameters. In all our calculations we have averaged over $10^3$ points in time, with two consecutive measurements separated by $10^4$ [$\sim N(\delta+\beta)$] Gillespie events. That is, between two measurements each particle is involved in one event on average. We find that $g_6$ decays exponentially with distance in the liquid phase, as a power law in the hexatic phase, and that it is approximately constant in the solid phase. 
These results are used to numerically characterize the different faces.
 A further measurement to confirm the phases (not shown) is the pair 
 correlation function, that only allows distinguishing 
 between the liquid and solid phases \cite{Caprini2019}.

As mentioned above and in contrast with conventional systems, the number of particles is not a set parameter in our model, and instead fluctuates in time in the stationary state.
One could therefore expect that the system moves from one phase to another as it undergoes these fluctuations. That is, without changing
any parameter, a phase transition would take place. We have not observed this in our model. Instead, the phase the system is in appears to be determined by the average total number of particles in the long-run. 

Despite the non-equilibrium character due to birth and death events, 
the phases are found to be homogeneous in space. In Fig.~\ref{Fig:Plot}(k)
we show the distribution of the local packing fraction across different positions in the system. In all three phases the distribution is unimodal.

\begin{figure}[hbt!]
\begin{center}
\includegraphics[scale=0.35]{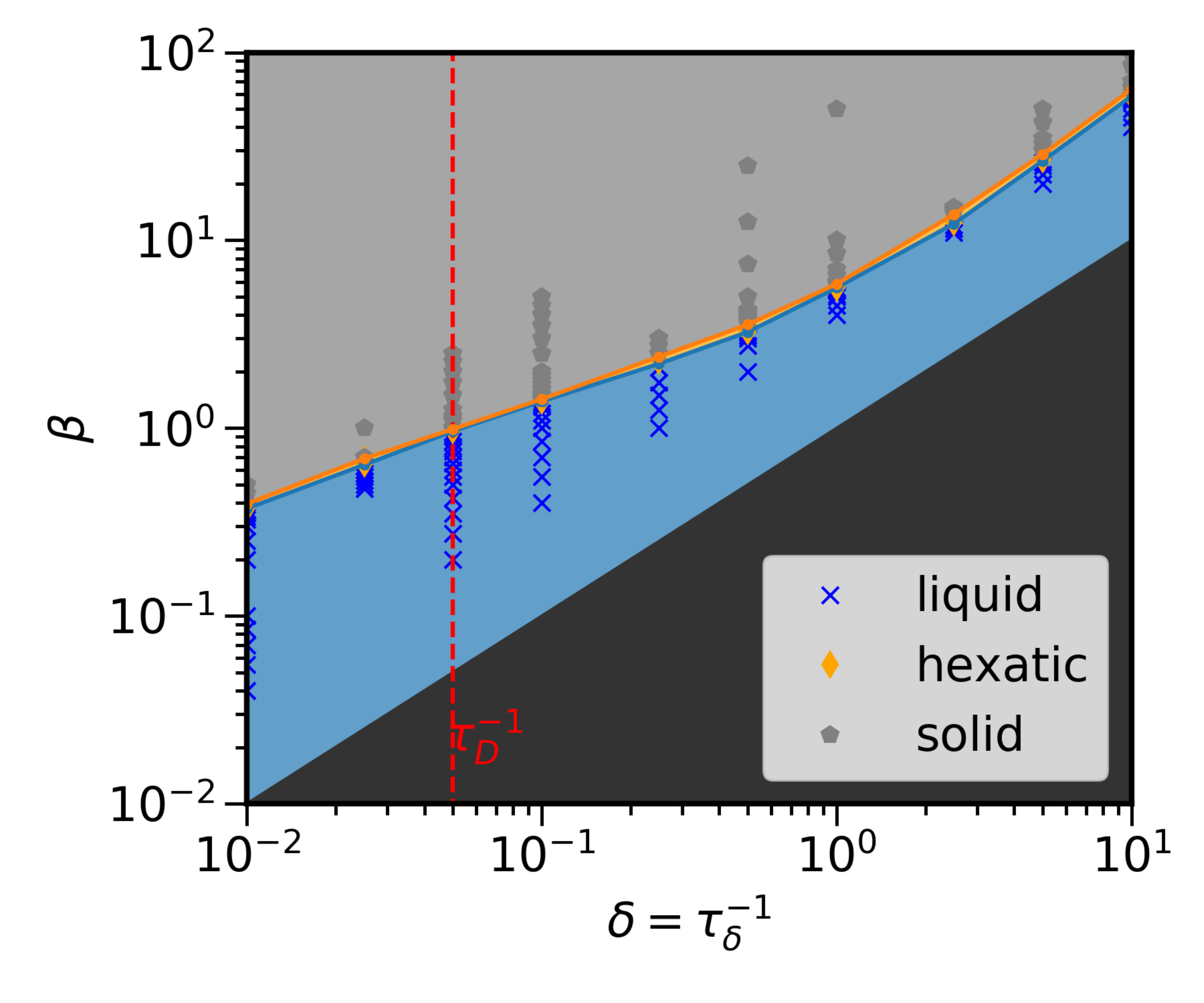}
    \caption{Phase diagram for passive Brownian particles with birth and death dynamics in the space spanned by the birth and death parameters ($\beta$ and $\delta$). The background colors represent the different phases (grey: solid phase, orange: hexatic phase, blue: liquid phase). The black 
      area represents the region of parameter space in which $\delta > \beta $. Markers show simulations for different choices of $\beta$ and $\delta$, with the type of marker indicating the resulting phase (see legend in the figure). The vertical dashed line marks the inverse of the diffusion time scale $\tau_{D}=\sigma^2/D$. }
    \label{Fig:phadiag}
\end{center}
\end{figure}

\subsubsection{Phase diagram}
Fig.~\ref{Fig:phadiag} shows the phase diagram of the system  in the plane spanned by the birth and death parameters, $\beta$ and $\delta$. We re-iterate that the density of particles depends on the birth and death rates, and is therefore not an independent control parameter as in conventional Brownian particle systems. 

For $\beta \gg \delta$ we find the system in the solid phase. In this situation the stationary filling fraction is high. When the birth parameter is comparable to the death parameter (but still maintaining $\beta>\delta$), the liquid phase is obtained. In a narrow range of parameters between these scenarios the system shows the hexatic phase. We have confirmed in simulations that the general features of this phase diagram are maintained as we vary the size $L$ of the system. 

The system has several natural time scales. One is the typical time between deaths, $\tau_{\delta}=1/\delta$, and another the time scale associated with diffusion, $\tau_{D}=\sigma^2/D$. The time scale associated with birth events changes as the number of particles in the system changes, and, in the stationary state, becomes equal to the typical time between deaths. In our simulations we have not observed any substantial difference in behavior in the two regimes $\tau_D > \tau_\delta$ and $\tau_D < \tau_\delta$.


\begin{figure}[hbt!]
\begin{center}
\includegraphics[scale=0.3]{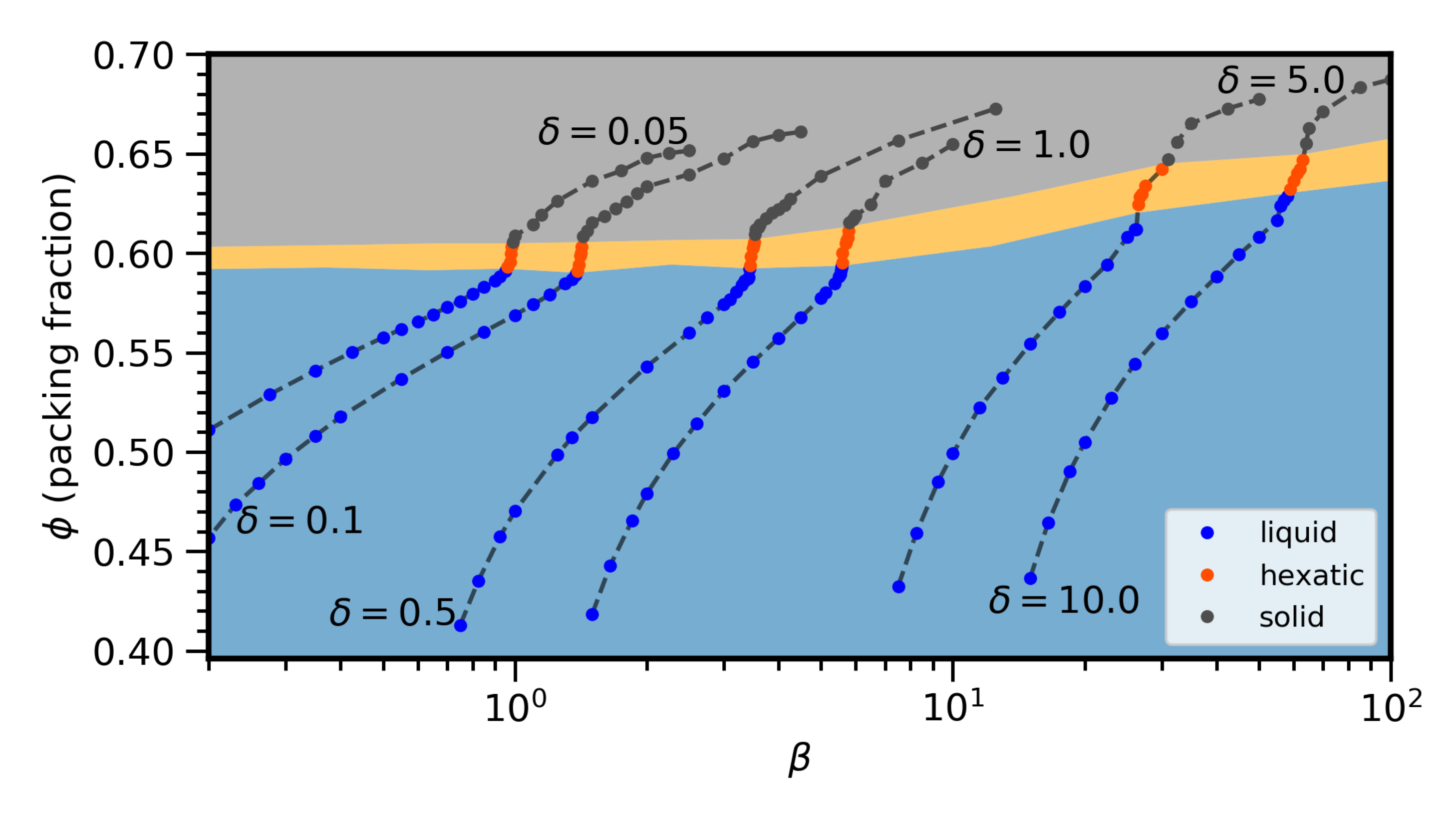}
    \caption{Packing fraction as a function of $\beta$ for different fixed values of $\delta$. Results from simulations are shown as markers. The background colors represent the three different phases (solid phase in grey, hexatic  orange, liquid phase blue).}
    \label{Fig:Nvslambda}
\end{center}
\end{figure}

\subsubsection{Packing fraction and number of defects}
In Fig.~\ref{Fig:Nvslambda} we plot the long-term packing fraction in the system as a function of the birth parameter $\beta$ for different fixed values 
of the death rate $\delta$. The background color indicates the three different phases, as in Fig.~\ref{Fig:phadiag}. Three regimes associated with the three different phases can be clearly identified. In particular, in the hexatic phase (simulations shown as red circles) the increase of the number of particles with $\beta$  is rather sharp. We also find that the packing fraction required for the system to be in the solid phase increases with increasing death rate $\delta$. This constitutes a difference
 with conventional PBP (at fixed temperature and without birth and death), where the packing fraction alone determines when the solid phase is observed.
 
\begin{figure}[hbt!]
\begin{center}
\includegraphics[scale=0.3]{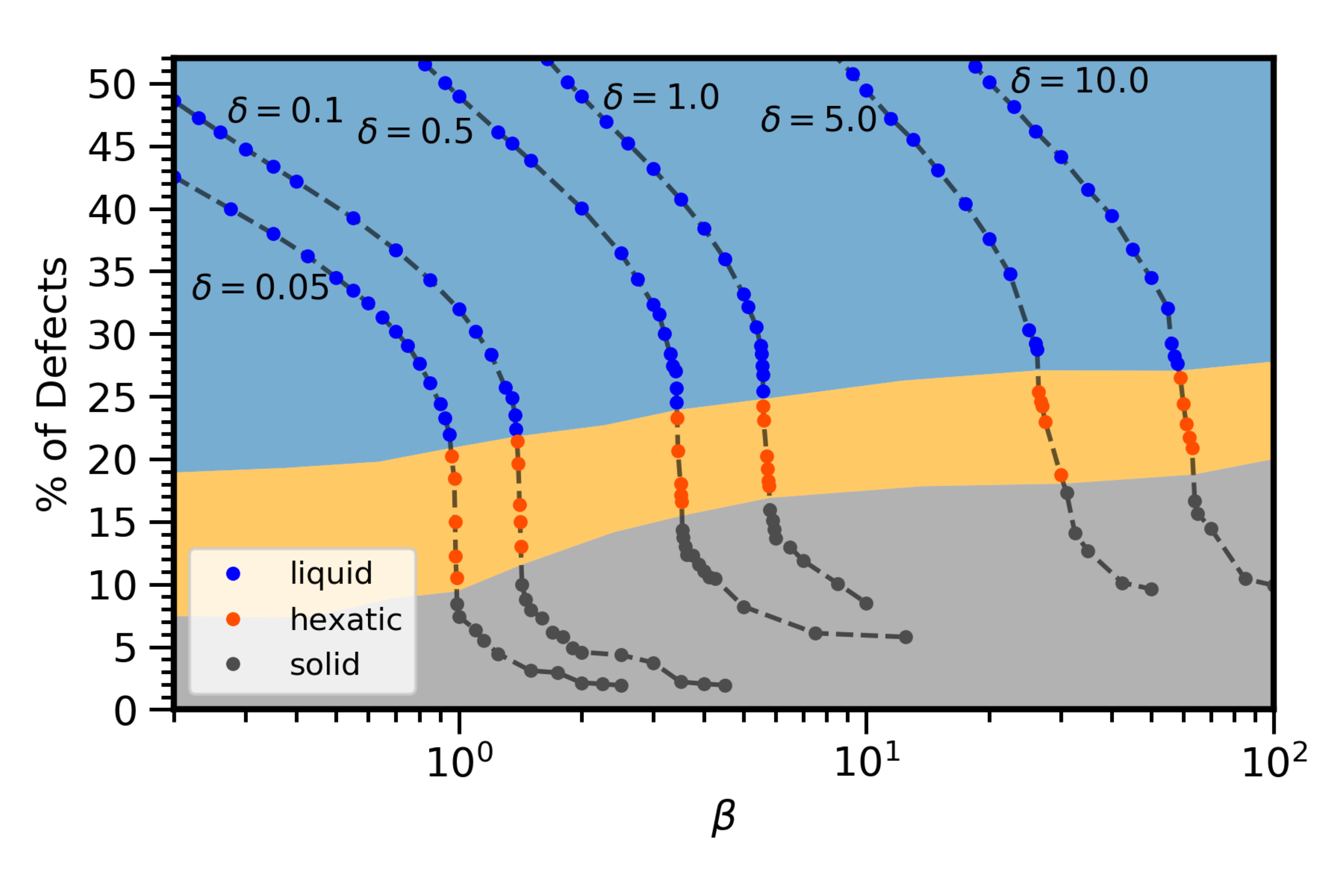}
    \caption{Percentage of defects as a function of $\beta$ for different values of $\delta$. Each curve represents a series of simulations with the same value of $\delta$ as indicated. The background colors represent the three different phases as in the two previous figures.}
    \label{Fig:defects}
\end{center}
\end{figure}

In Fig.~\ref{Fig:defects} we show the percentage of
defects (particles that do not have exactly six neighbors)
 as a function of the birth parameter $\beta$ for 
different fixed values of the death rate $\delta$. 
  The data in Fig.~\ref{Fig:defects} shows strong dependence 
on the birth and death parameters. We also find that the 
transition between the hexatic and solid phases occurs
at a fraction of defects of approximately $10\%$ to $25\%$. 
 This is significantly higher than the percentage of defects 
of approximately $5\%$ observed in
 standard PBP and ABP systems at this transition
\cite{digregorio2018,Digregorio2022}.

\subsection{Active particles}
\label{subsec:active}

We next study the ABP system with birth and death events. We
 fix the death rate to $\delta=0.01$. When a birth event 
 occurs then the propulsion force of the offspring 
is initially set to be in same direction as that of the
 parent particle. Alternatively, the direction of the propulsion force of the offspring can also be chosen at random. Simulations (not shown) indicate that the main properties of the system remain unchanged.
 
\begin{figure}[hbt!]
\begin{center}
\includegraphics[scale=0.42]{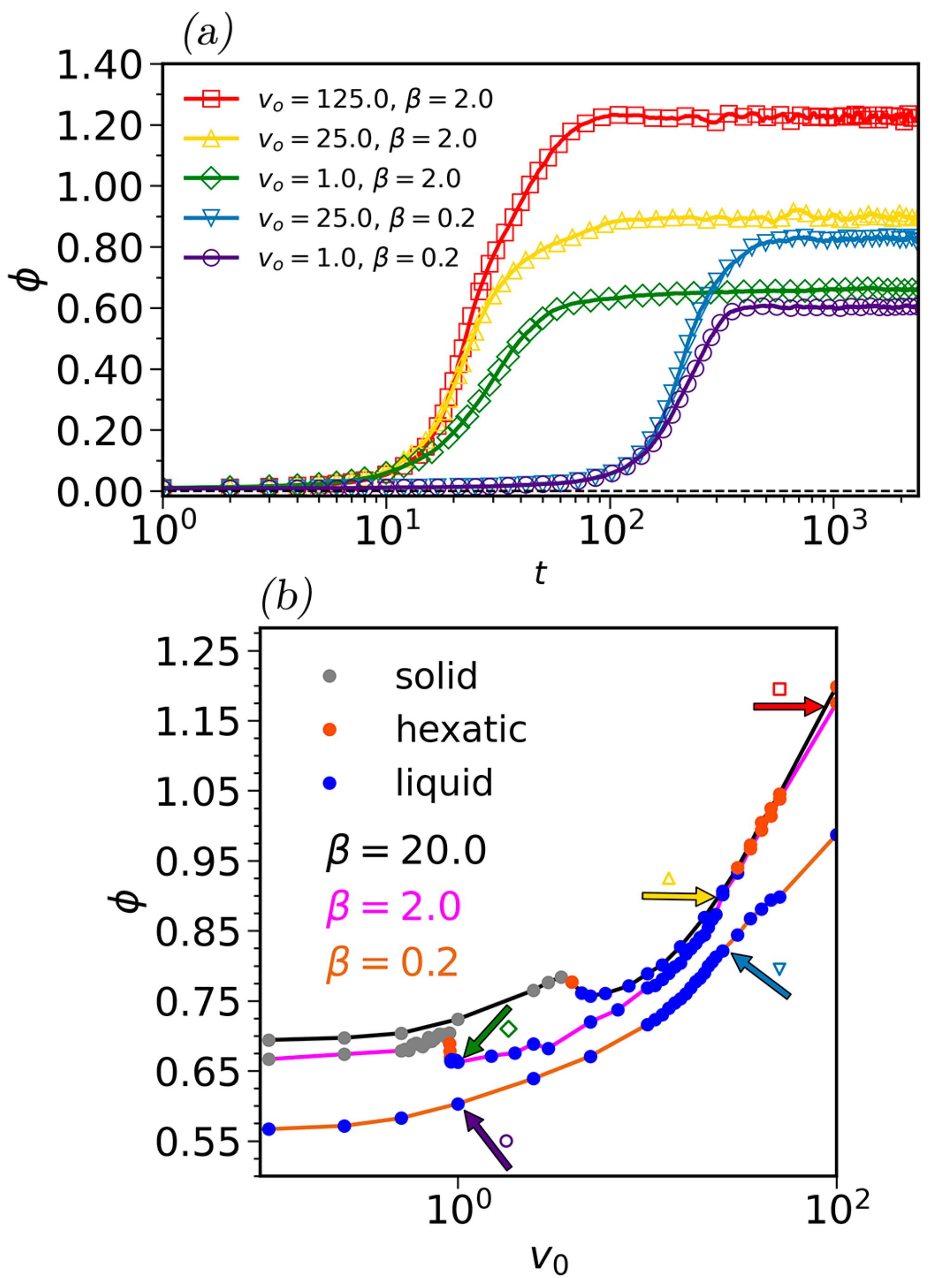}
    \caption{(a) Packing fraction versus time for different
values of $\beta$ and strength of the active force ($v_0$) for $\delta=0.01$. 
(b) Long-time particle fraction versus 
$v_0$ for different values of $\beta$.
The colored arrows indicate the parameter choices used in panel (a).} 
    
    \label{Fig:Nvstactive}
\end{center}
\end{figure}
 \subsubsection{Packing fraction}
Fig.~\ref{Fig:Nvstactive}~(a) shows how the packing fraction in the system evolves in time, for different values of the birth and death rates, and of the activity parameter $v_0$. As in the passive case, when the value of $\beta$ is increased, keeping all other parameters fixed, the packing fraction also increases. On the other hand, an increased activity typically leads to a higher number of particles in the system at long times, in particular the packing fraction in the active system is higher than that for passive particles. This is further detailed in Fig.~\ref{Fig:Nvstactive}~(b) for a selection of fixed values of the birth and death parameters. We find that the steady-state packing fraction only 
experiences a minor increase with the activity when $v_0$ is small, 
but that the increase is much more pronounced when $v_0$ is large (in other words, the packing fraction curves upwards as a function of the activity). The reason for 
the increase is that when the activity is high enough the
 particles may penetrate the repulsive potential and overlap (at these values the average distance between particles 
gets smaller than $\sigma$). Then there is more room for self-replication, and 
the steady-state number of particles increases. 
For intermediate values of $v_0$, the packing fraction 
diminishes signalling the transition from solid to liquid. 
Overall, we conclude that both the birth/death rates and the activity determine the long-time packing fraction of active particles and thus the structural phases of the system. 

\subsubsection{Motility-induced phase separation and phase diagram}
\begin{figure}[hbt!]
\begin{center}
\includegraphics[scale=0.47]{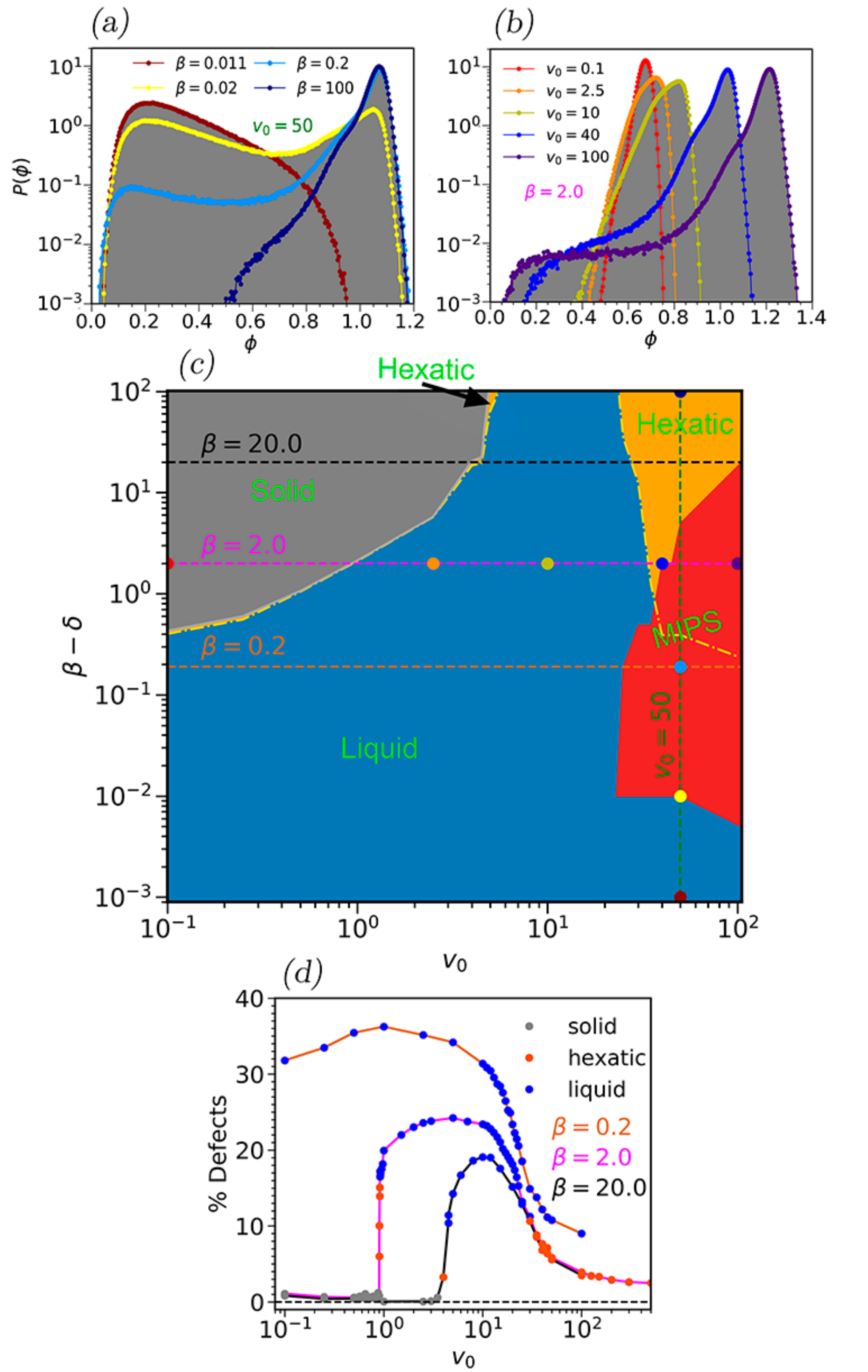}
\caption{
Panels (a) and (b): Distribution of the local packing fraction for different values of the birth parameter $\beta$ and the activity parameter $v_0$. In panel (a) we fix $v_0=50$ and vary $\beta$, in panel (b) $\beta=2$ is fixed and $v_0$ is varied.
Panel (c): Phase diagram for the ABP system.
Background color in panel (c) represents the different phases (solid phase grey, hexatic orange, 
liquid phase blue, MIPS red ). The different phases are also indicated.
The parameters used in panels(a) and (b) are indicated by colored markers in panel (c). Panel (d) shows the density of defects as function of $v_0$ for $\beta = 0.2, 2, 20$ 
 [corresponding to the three horizontal lines in panel (b)].}
\label{Fig:MIPS}
\end{center}
\end{figure}

In addition, the activity has a similar impact on the system  as in standard ABP, that is, we observe 
nonhomogeneous phases or motility-induced phase separation. That is to say, we find a separation of dilute and dense phases for sufficiently high self-propulsion.
This is shown in Fig.~\ref{Fig:MIPS}~(a) and (b) where the MIPS phase is identified from the bimodality 
of the local packing fraction distribution, $P(\phi)$, at long times. The two maxima represent the dilute and dense phases.


Fig.~\ref{Fig:MIPS}(c) shows the phase diagram of the system (in terms of $\beta$ and $v_0$, for fixed $\delta$).
We observe that for $v_0\lesssim 5$ the system behaves broadly as in the absence of activity. The effects of activity on the packing fraction are
small in this regime [see Fig.~\ref{Fig:Nvstactive}(b)], and thermal fluctuations dominate. Since self-propulsion plays the role of an effective temperature \cite{Levis2015,Palacci2010}, the larger $v_0$ the liquid-hexatic and hexatic-solid transitions are shifted towards larger density (or lower temperature). In our case, this translates into a larger birth rate.
 
For large activity we find interesting differences in the phase diagram of our system [Fig.~\ref{Fig:MIPS}~(c)] compared to that of the standard ABP system without birth and death (this phase diagram is shown for example in Fig.~1 of \cite{digregorio2018}). The most important difference is a re-entrance phenomenon, seen  for example along the cuts $\beta=2.0$ and $\beta=20.0$, indicated by the horizontal lines in Fig.~\ref{Fig:MIPS}~(c). As the activity parameter is increased the system is first in the solid phase, then hexatic, and then liquid, and then, upon further increase of the activity, it becomes ordered again and returns to the hexatic phase. 

To investigate this further we plot in Fig.~\ref{Fig:MIPS}(d) the density of defects versus $v_0$ along the cuts $\beta=0.2, 2.0, 20.0$. The percentage of defects is non-monotonic and reaches a maximum at intermediate values of the activity $v_0$. We note that this is very different to what happens in standard ABP systems where the activity acts as an effective temperature, so that the system becomes more disordered with increased activity, resulting in an increased percentage of defects. 
In the system with birth and death, larger activity induces particle aggregation, leading to clusters that are quite ordered due to their high density. In an active system without population dynamics, these particle clusters are also formed as the activity is increased, but free space remains and the system does not become ordered. In the system with birth and death, production events fill this available space with additional particles. The resulting clusters pervade the entire spatial domain for large activities, so that the system becomes ordered, and returns to the hexatic phase. In this regime we note that particles get closer to each other due to the increased activity. This can be seen in Fig.~\ref{Fig:MIPS}(b), where the peak of the distribution of local packing fractions shifts to the right, corresponding to a larger packing fraction. 

Similar effects can be seen upon increasing the birth rate, while keeping $v_0$ fixed at sufficiently large values to have a MIPS regime. If the birth rate $\beta$ is small  (e.g. $\beta<0.02$ for $v_0=50$), the system is in the liquid phase. As $\beta$ increases, an increased number of birth events aggregates
particles into denser local clusters in some regions, thus producing MIPS. As $\beta$ is increased further, these denser regions grow and occupy more of the system. MIPS is then no longer found, and the system becomes hexatic. This is also confirmed in Fig.~ \ref{Fig:MIPS}~(a).
As the birth parameter $\beta$ is increased the maximum of the distribution of the local packing fraction at $\phi \simeq 0.2$ reduced in height. In contrast, the maximum at $\phi \simeq 1.1$ is then the only one that persists. For intermediate values one observes MIPS, and both maxima of $P(\phi$) are present. 

\section{Summary and discussions}
\label{Sec:summary}

As a minimal model for the biological aggregation of finite-size moving individuals, we have studied systems of passive or active Brownian particles in which particles may also die and self-replicate at given rates. To prevent an absorbing state of a system void of particles, we have focused on scenarios in which birth dominates over removal. Both in the system with active and with passive particles the long-time packing fraction depends on the birth and death parameters, and these thus determine what structural phases are observed. In the case of active particles, the stationary particle density is additionally affected by the strength of self-propulsion. In simulations we have constructed the phase diagram in terms of  the birth and death parameters for the passive system, and in terms of birth parameter and activity in the active system. Liquid, hexatic and solid phases are observed, and additionally phase coexistence (MIPS) in the active system.

A fundamental feature of our system is that, due to the birth and death dynamics, the number of defects is larger than in conventional systems. At difference with the standard ABP system, we find the hexatic phase at large activity (there is re-entrance to this phase), reflecting the increased number of defects due to the combination of birth, death and activity.

Studying the dynamics of particle systems combining motion and particle proliferation is relevant for a number of biological processes.
For example, wound healing and tissue formation have been studied using models of interacting particles \cite{Zapperi,Sepulveda}. Analyses such as ours can then contribute to gaining a deeper insight into the mechanics of these systems.
Our approach is complementary to existing models without birth and death processes. It provides qualitative information about the spatial ordering of cells progressing through an irregular border, or about optimal demographic rates to improve the average progression.


Future extensions of this work could consider 
more complex biological ingredients, like 
diversity for the size and shape of the particles, 
other interactions, confinement or 
competition for resources,
but also some other types of movement
such as run and tumble dynamics and Levy-like flights \cite{Huda2018},
and different descriptions of the activity \cite{CapriniSciRep}.

\section*{Acknowledgements}

We acknowledge Lorenzo Caprini and Emilio Hern\'andez-Garc\'\i a 
for a critical reading of the manuscript.
We also
acknowledge
MCIN/
AEI/10.13039/501100011033/ and FEDER “Una
manera de hacer Europa” for its support to the project
MDM-2017-071, Maria de Maeztu Program for Units
of Excellence in R$\&$D.


\nocite{*}
\bibliography{bibl.bib}

\end{document}